\documentclass[runningheads]{llncs}
\usepackage{graphicx}
\usepackage{tabularx}
\usepackage{booktabs}
\usepackage{comment}
\begin{document}
\title{Quantifying Population Exposure to Long-term PM\textsubscript{10}: A City-wide Agent-based Assessment\thanks{Supported by MRC/CSO Social and Public Health Sciences Unit}}
\titlerunning{Quantifying Population Exposure to Long-term PM\textsubscript{10} using ABM}
%
\author{Hyesop Shin\inst{1}\orcidID{0000-0003-2637-7933}}
\authorrunning{Hyesop Shin}
%
\institute{MRC/CSO Social and Public Health Sciences Unit, School of Health and Wellbeing, University of Glasgow, Clarice Pears Building, 90 Byres Road, Glasgow, G12 8TB, UK\\
\email{hyesop.shin@glasgow.ac.uk}}
%
\maketitle              
\begin{abstract}
This study evaluates the health effects of long-term exposure to PM\textsubscript{10} in Seoul. Building on the preliminary model \cite{shin2019}, an \textit{in-silico} agent-based model (ABM) is used to simulate the travel patterns of individuals according to their origins and destinations. During the simulation, each person, with their inherent socio-economic attributes and allocated origin and destination location, is assumed to commute to and from the same places for 10 consecutive years. 
A nominal measure of their health is set to decrease whenever the concentration of PM\textsubscript{10} exceeds the national standard. 
Sensitivity analysis on calibrated parameters reveals increased vulnerability among certain demographic groups, particularly those aged over 65 and under 15, with a significant health decline associated with road proximity. 
The study reveals a substantial health disparity after 7,000 simulation ticks (equivalent to 10 years), especially under scenarios of a 3\% annual increase in pollution levels. 
Long-term exposure to PM\textsubscript{10} has a significant impact on health vulnerabilities, despite initial resilience being minimal. 
The study emphasises the importance of future research that takes into account different pollution thresholds as well as more detailed models of population dynamics and pollution generation in order to better understand and mitigate the health effects of air pollution on diverse urban populations.

\end{abstract}
\section{Introduction}
\label{sec:Introduction}

Exposure to air pollution has been extensively studied in order to comprehend how health consequences of air pollution on populations can vary greatly \cite{Guarnieri2014}. According to some studies, exposure levels vary depending on one's socioeconomic profile, i.e. the young and old, the socially disadvantaged, and those living near polluted areas have higher exposure and health risk \cite{ONeill2003,Kan2004,Kan2008,David2012,Min2020}. 
Air Pollution and Inequalities in London 2019 \cite{tim2021} stated that the most deprived communities in London were 13\% and 6\% more likely, respectively, to be exposed to higher levels of NO\textsubscript{2} and PM\textsubscript{2.5} than the least deprived areas. Although annual concentrations of these pollutants and population exposure in London have decreased since 2013, there are still disparities in exposure in more deprived areas, as well as across age and ethnic groups.

To investigate the association between air pollution, socioeconomic status, and health, both top-down approaches, such as spatial interpolation and dispersion models, and bottom-up approaches using sensors and GPS gadgets were used. 
While the top-down methods had the advantage of indicating a population-wide exposure measure, the limitation is that the exposures were measured with an aggregated figure (annual mean of PM\textsubscript{10}), vaguely assuming that the population's exposure happened at their home locations \cite{Wong2004,David2012,Beevers2013,Min2020,CambridgeshireCityCouncil2016,Nyhan2016,Beevers2016}.
Studies using bottom-up methods, on the other hand, have significantly discovered more sophisticated exposure levels by tracking people's mobility, but the small number of participants, time-consuming recruiting, and short-term modelling period were insufficient to generalise the findings\cite{Steinle2015,HWANG2018192,Liang2019,Sanchez2020,Larkin2017}. 
Thus, despite advances in tools and data, the translation between air pollution, exposure, and health effects across individuals has remained highly uncertain.

Agent-based modelling (ABM) is one of the methods to generate an individual’s unique attributes (e.g. age, gender, area of residence), mobility patterns, and association with air pollution. Typically, the individuals are situated in a confined space and time and move across tessellated or networked neighbourhoods according to the geospatial structure. This can advantage enable us to quantify the cumulative exposure to air pollution of each individual and the adverse health consequences.

This paper investigates the cumulative effects on population exposure to PM\textsubscript{10}, individual's socioeconomic backgrounds, and the mobility patterns of population health. The specific research questions are:
   \begin{itemize}
   \setlength\itemsep{0em}
     \item How can personal exposure translate to a health risk?
     \item How do demographic background and mobility patterns potentially affect health outcomes? 
     \item How could the population health outcome differ by future pollution scenarios? 
   \end{itemize}

Building upon the proof-of-concept model made by Shin and Bithell \cite{shin2019}, this study expanded an \textit{in-silico} agent-based model (ABM) to all 25 districts of Seoul.

\section{Methods}
\subsection{Data Collection}\label{section:datacollection}

\subsubsection*{Pollutants}
A series of hourly PM\textsubscript{10} was collected from the nearest background stations from each district between 1 January 2010 and 31 December 2015, then was grouped by home hours (assumed to be 20:00–08:00) and working hours (09:00–19:00). 
To account for some missing observations in the data (e.g. 2.15\% in Gangnam and 785 hours due to periodical inspections), we used a Kalman algorithm to fill the missing values using an \texttt{ImputeTS} package in R \cite{Moritz2017}.

\subsubsection*{Demographics}
Age was included, as health risks to air pollution can depend on personal physical condition, and in general, are prominent in vulnerable age groups \cite{Pearce2006,Pearce2011}. 
The 2010 population data of Seoul by a five-year interval were grouped as 5-9, 10-14, …, 80-84, and over 85. 

\subsubsection*{Land Price}
\begin{figure}[h]
 \begin{center} 
	\includegraphics[width=.9\textwidth]{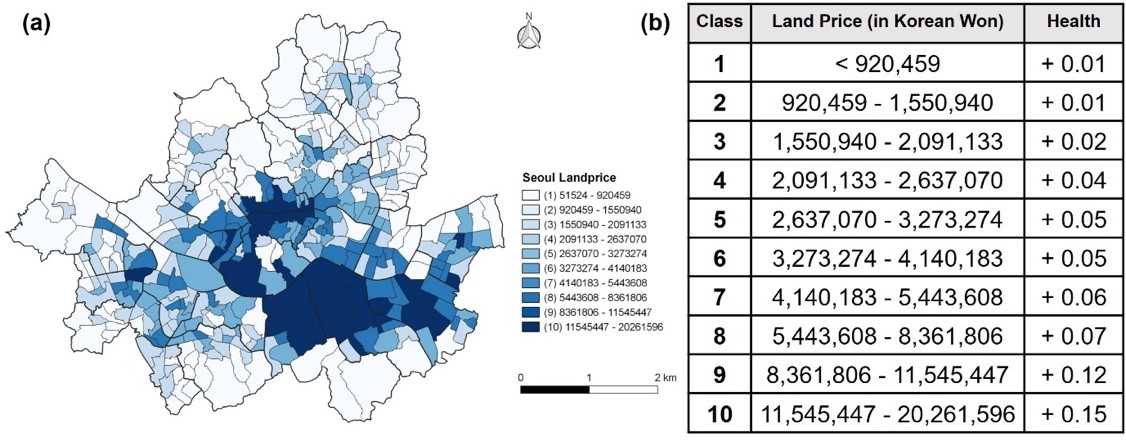}
 \end{center} 
 \vspace{-5mm}
  \caption{Official land prices in Seoul sub-districts in 2015 (a), and table of hypothetical health changes determined by land price in Korean currency (\$1$\approx$₩1,200) (b)} \label{fig:landprice} 
 \end{figure}

The Official Land Price was selected as a proxy for the rate of recovery from the effects of air pollution (see Figure~\ref{fig:landprice}). Residential property was chosen not only because it represents immovable and location-specific capital, but also because it expresses price in the economy: those who can afford higher-priced housing may be better able to access health care or adjust their lifestyle to compensate for high pollution levels. 


\subsubsection*{Origin-Destination}

Origin-destination matrices for Seoul were used to infer the population's home and work locations. 
Given the fact that OD matrices produce a coarse temporal resolution, namely two locations per day, the study assumed that the population moves across the land cover of either Residential, Commercial, or Traffic areas that would be likely for home and work locations.

\subsection{Agents}
\subsubsection*{Population Sampling} We used a 5\% population sample of agents to generate a simple synthetic population in Seoul districts. Each agent has a list of attributes including home location, work location, age group (i.e. young, active, or old), and health. 

\subsubsection*{Nominal Health} Since individual health records or exposure histories were unavailable, agents were assumed to have had no prior exposure. As a result, our simulation only simulates the likely rate at which agents accumulate exposure effects over time in a given district. Each agent is given a `nominal health' level, which is an integer with a value of 300 at the start.
Depending on their socioeconomic status, they will incur a decline in health if exposed to PM\textsubscript{10} levels above a threshold near 100µg/m\textsuperscript{3}, which is South Korea's hourly air quality standard. When the ambient pollution level exceeds the standard, the health loss functions will activate.

\subsubsection*{Agent Behaviour} 
\begin{figure}[hbt!]
 \begin{center} 
	\includegraphics[width=.8\textwidth]{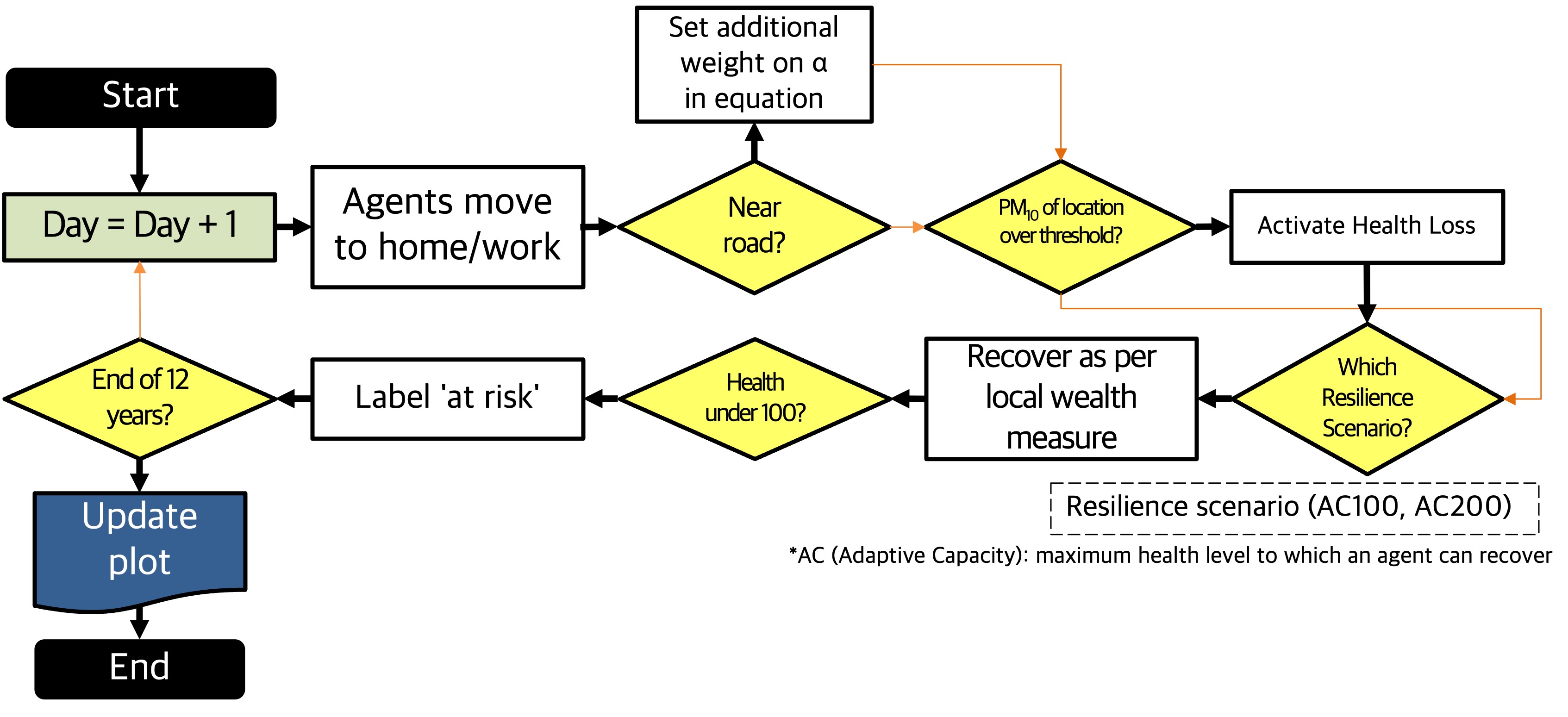}
 \end{center} 
 \vspace{-5mm}
\caption{The implementation algorithm for each period. Retrieved and edited from Shin and Bithell \cite{shin2019}} \label{fig:algorithm} 
 \end{figure}

The model's agents have limited reasoning abilities (see Figure~\ref{fig:algorithm}). Their behaviour are entirely driven by a simple timetable. All agents adhere to the following hypotheses:

\begin{itemize}
  \itemsep0em 
  \item An agent’s birth, death, and ageing are not considered
  \item Agents understand their origin (home) and destined (work/school/outdoors) locations but have no cognitive perception
  \item 1 tick is equivalent to half a day, that is, Work hours (09-19 hrs), Home hours (20-08 hrs)
  \item Every agent starts with a health status of 300, but this drops when the agent is exposed to PM\textsubscript{10} over 100µg/m\textsuperscript{3}
  \item Agents move from and to either residential, commercial, or traffic areas
  \item Agents commute to the same location until the simulation ends
  \item For visualisation purposes, if the health status of an agent drops below 200, the agent’s colour turns purple; when it drops below 100, the colour turns red
  \item If an agent’s health reaches 0, the agent will be sent to the hospital for treatment
  \item All agents stop if the system reaches 8764 ticks (equivalent to 12 years) or if the `at risk’ population reaches 100\% of the total population
\end{itemize}
 
\subsubsection*{Movement} Agents between the ages of 15 and 64 commute within a sub-district but can also move to different sub-districts based on the origin-destination matrix. Agents under the age of 15 will move to a random patch within radius 3, while those over the age of 65 will move to a random patch within radius 1: this is intended to represent a more limited range of movement for this demographic group. We simplified movement by translating the agents to their destination patch during the day (one tick) and back to the home patch at night (the next tick) because the traffic flow is not considered. Agents who travel within the district are assessed, whereas those who travel beyond the boundary are excluded. 

\subsection{Quantifying Personal Exposure and Health Effects} \label{exposure-healtheffects}

In general, long-term pollution exposure and health outcomes are non-communicable \cite{fatmi2020air}. As a result, the model only considers the effect of continuous interactions between individual spatial trajectories constrained by daily activity patterns and the atmospheric pollution distributions assigned to each latticed location.

This study used the equation of Shin and Bithell \cite{shin2019} that set the rate of change of an individual’s health status caused by PM\textsubscript{10} exposure to varying (non-)linearly with health:

\begin{equation}
\label{eq:healthloss}
  If PM\textsubscript{10} \geq 100, \qquad
  \frac{dH}{dT} = -\alpha\eta(H\textsubscript{max} - H(t)) + H\textsubscript{recov}
\end{equation}

\begin{itemize}
  \itemsep0em 
  \item H\textsubscript{max}: an agent’s health status at the beginning
  \item H(t): current (\texttt{t}) health value
  \item $\alpha$: the rate of change per unit of time when the health impact applies (0$<$$\alpha$$<$1). Agents’ health values would decrease exponentially away from their initial value H(0).
  \item $\eta$: An additional health loss parameter that is age dependent (same as the $\alpha$ effect but tweaked after validation)
  \item H\textsubscript{recov}: a health recovery rate that varies by the real estate price of the agent’s home location, as in Figure ~\ref{fig:landprice}, up to maximum adaptive capacity (AC, i.e. recovery can only increase health up to some scenario-dependent value).
  \item Road patches: a 1.5-fold of background PM\textsubscript{10} was added to the road patches to elevate the particulate levels near road patches. As with the health loss parameter, parameters for road pollution were also tested in the following section.
\end{itemize}

The agent exposed to PM\textsubscript{10} above the threshold will steadily lose its health. However, compared to an agent with a health of 250, an agent with a health of 80 will lose health more rapidly when they are equally exposed to over 100µg/m\textsuperscript{3} of PM\textsubscript{10}.

\subsubsection*{Health Outcome}
This study coins the term ``at-risk'' population to assess the overall health outcome.
An agent who is ``at-risk'' means the individual's nominal health is below 100 (i.e. a third of the initial health condition). The ``at-risk'' population means the number of people whose nominal health is below 100, and the at-risk rate is computed from the following equation: \textit{``at-risk" population} / \textit{population}.
This study uses at-risk, health risk, health outcome, and the population at-risk interchangeably.

\section{Model Interface} \label{Model interface}
\begin{figure}[h]
\begin{center} 
	\includegraphics[width=.7\textwidth]{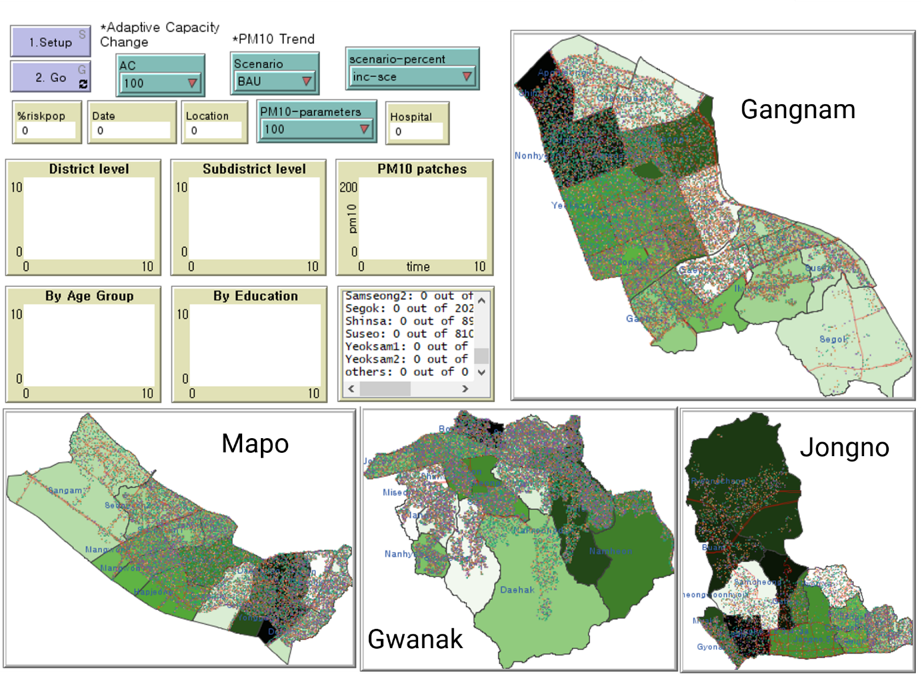}
\end{center} 
\vspace{-5mm}
 \caption{Implementation in Netlogo 6.0.4 for Gangnam, Mapo, Gwanak, and Jongno} \label{fig:netlogo} 
\end{figure}

The model environment was derived from a GIS data set of each district (see Figure~\ref{fig:netlogo}). 
For simplicity, building and traffic information was excluded for the present. The spatial extent of Gangnam, Mapo, Gwanak, and Jongno, below, are around 40km\textsuperscript{2}, 24km\textsuperscript{2}, 30km\textsuperscript{2}, 24km\textsuperscript{2}, respectively, with a 30m × 30m spatial resolution. This study simulated 12 years using two time-steps per day (home hours and working hours), with the earlier six years of PM\textsubscript{10} coming from the observational data set. The first six years for which data was available were reused but modified to create scenarios for future projections for the latter six years.

\vspace{-5mm}

\vspace{-10mm}

\section{Sensitivity and Calibration} \label{section:Sensitivity and Calibration}

We assessed the main factors that distinguish the size of the total at-risk population and then calibrated the population with hospitalised patients. For each parameter, a one-factor-at-a-time (OFAT) analysis was used.

\subsection{Measuring Sensitivity to the Risk Population}

Two parameters were included for the sensitivity analysis: health loss and road proximity. 
Gwanak and Gangnam were chosen for their similarity in area size, but also for the distinctiveness of land prices (recovery rate), as they represent one of Seoul's most and least deprived districts, respectively.
For each implementation, the study plots time on the x-axis and the proportion of the population at risk on the y-axis. To reduce stochastic variability, the outcome was averaged over 20 iterations. It is important to note that the stochastic variability was negligible.

Table~\ref{table:calibration} is a summary of the parameters selected and tested. 
Health-loss (general) was used to control the leverage of health loss across the entire population, whereas Health-loss (age groups) was used to adjust the values by age group, but is not shown here because each age group used different parameters to fit the hospital admission data.
Road proximity was tested from 1 to 1.5 and 2 times greater PM\textsubscript{10} concentration than the background, assuming no additional effect from the roads.

\begin{table}
\caption{Demographic parameter values used for calibration and tested for model sensitivity}
\label{table:calibration}
\centering
\begin{tabular}{l l l}
\toprule
Parameters & Selected & Tested \\
\midrule
Health loss (general) & 0.0043  & 0.001 - 0.01 (by .001) \\
\midrule
Health loss (age groups) & Differs by age groups  & Numerous combinations \\
\midrule
                      &      & 1     \\
Road proximity        & 1.5  & 1.5   \\
                      &      & 2     \\
\bottomrule
\end{tabular}
\end{table}

\begin{figure}[h]
\begin{center} 
	\includegraphics[width=.55\textwidth]{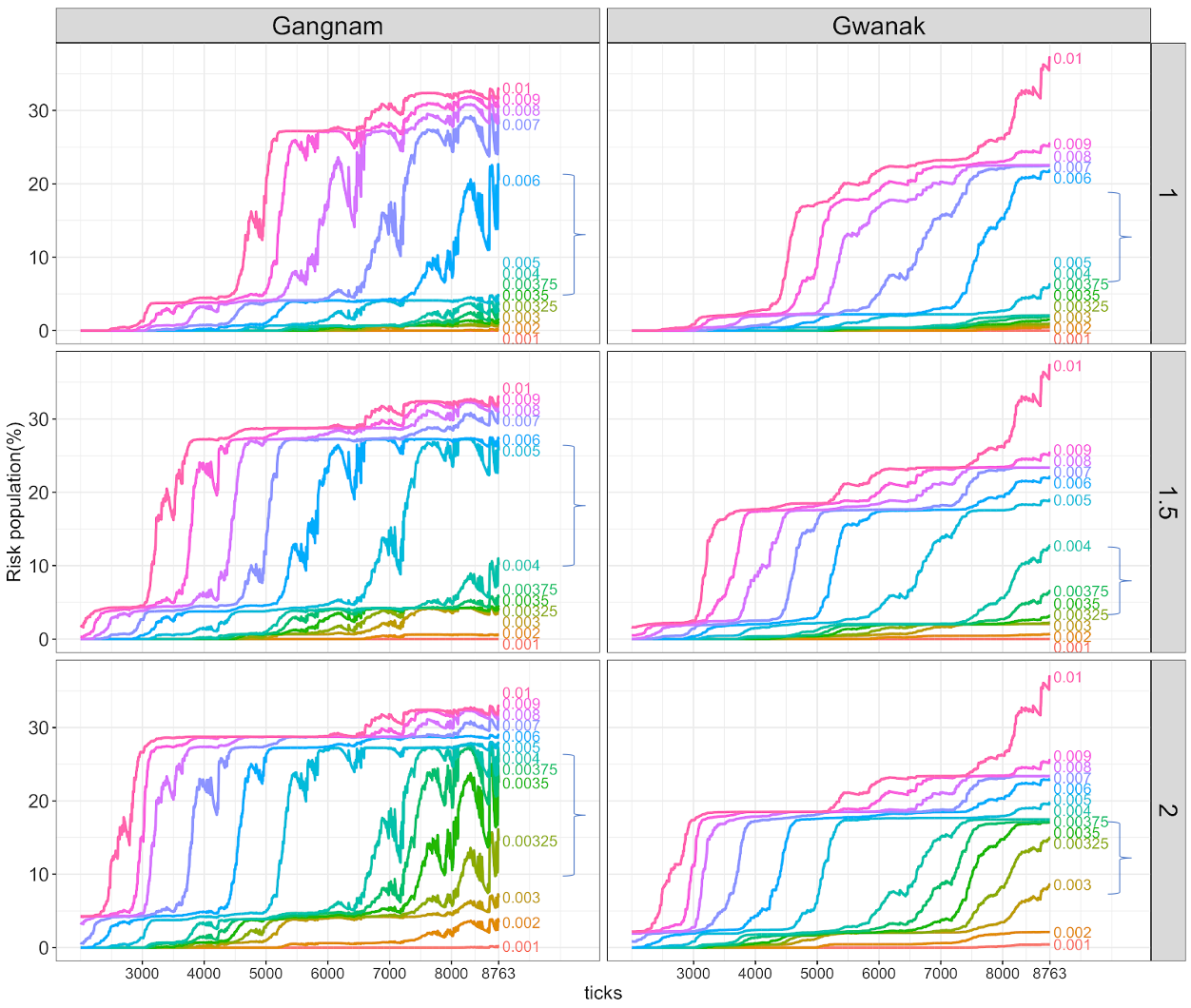}
\end{center} 
\vspace{-5mm}
\caption{Sensitivity of health loss and road proximity by time and risk population. The 3X2 array is indexed by districts and pollution weights to road patches. The decision threshold is indicated by the curly brackets. The decision threshold in both districts is between 0.003-0.005.} 
\label{fig:sensitivity_line} 
\end{figure}

Figure~\ref{fig:sensitivity_line} illustrates the sensitivity of health loss against road proximity. Overall, health loss was sensitive to small unit changes, where a 0.001 rise of $\alpha$ can result in a 10-20\% difference in the health risk output. Adjusting the parameters from 0.003 to 0.2, the health risk of Gangnam and Gwanak resulted in 0-32\% and 0-37\%, and both districts showed a 16-18\% gap of at-risk rates between 0.005 and 0.006. Notably, the timing of health risk surge varied by districts but also the road proximity parameters. The tipping point between districts was mainly due to the 15-64 age group whose health went below 100. Road proximity affects the outcomes to change linearly as the parameters increased.

Having tested the combinations between both parameters and value adjustments, this study selected health loss at 0.0043 and road proximity at 1.5 to be implemented in scenarios, with the evidence that PM\textsubscript{10} on roads are approximately 50\% higher on average than the background concentrations. The final parameter value resulted from the calibration with the hospital admission data from HIRA (Health Insurance Review \& Assessment Service). The processes will be further explained in the next section. Note that the selected parameter values are all illustrative and it can depend on the adjustment of other settings, e.g. calibration data.


\subsection{Calibrating Unwell Agents to CDC Patient Data}
\subsubsection{Data Description}

The model counted those with a health score of 0 or lower as being in a hospital for treatment. The Korean CDC (Center for Disease Control and Prevention) database, known as HIRA, made hospital patient data available. The database extracted respiratory patients (sections J01-J99) from 2016, excluding J00 (common cold) and patients treated by medical herbalists (acupuncture, moxibustion and so on). The data was provided for non-commercial purposes at a cost of about \$300 per year.

Since each patient's home address was anonymised (legal privacy protection), this study summed patient statistics from all Seoul hospitals by age groups in five-year intervals, e.g. 5-9, 10-14. To match the outcome of the simulation that used 5\% of the census, the total number of 230,962 patients from each age group was reduced to a fraction of 5\%. Finally, patients from each district were added and compared to the final figure from the BAU-AC100 scenario. 


\subsubsection{Calibrated results}

\begin{figure}[h]
 \begin{center} 
	\includegraphics[width=.7\textwidth]{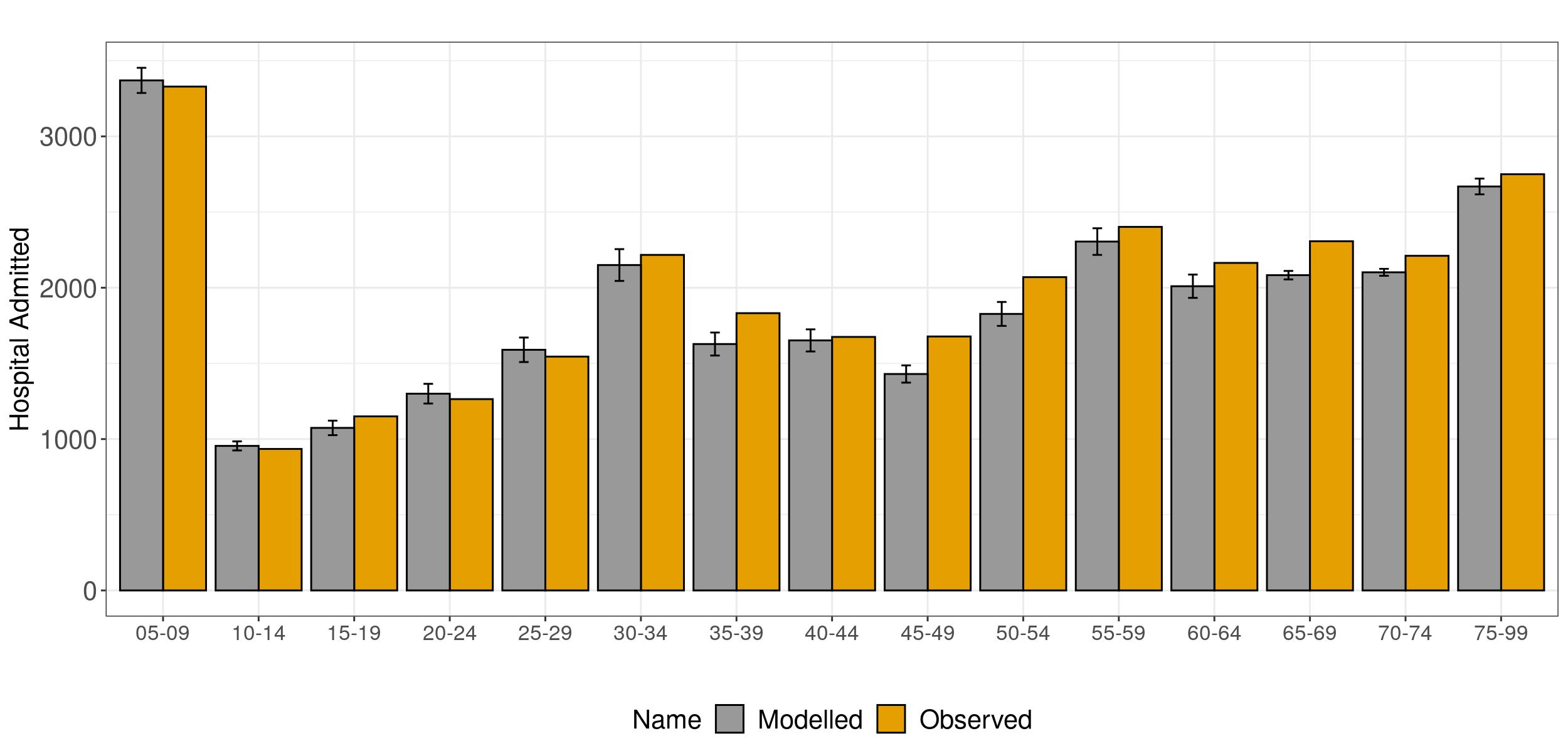}
 \end{center} 
 \vspace{-5mm}
 \caption{The modelled result was calibrated against patient data. The observation data is based on a two-year average of the Korean CDC from 2015 to 2016} \label{fig:calibrated} 
 \end{figure}

From 25 district outcomes of the BAU-AC100 scenario, the total number of patients from the model made a reasonable approximation to the observed data (see Figure~\ref{fig:calibrated}). 
The majority of the group had a difference of $<$100 between the modelled and observed values.
The difference between the model and the observation might be reduced to less than some percentage of the total, and the variance in the model runs was typical of a similar order of magnitude. 
The outcomes from the model run revealed that 11 out of 25 districts had patients admitted to the hospital\footnote{The selected districts include Dongdaemoon, Gangseo, Guro, Gwanak, Gwangjin, Jung, Jungnang, Mapo, Seocho, Yangcheon, and Yeongdeungpo, however, the majority of the patients were found from Gwanak and Mapo}.

\vspace{-10mm}

\section{Scenario Analysis}\label{section:results}
\subsection{Scenario Description}
Following the model’s calibration, this study set two `what-if' scenarios for pollution trends and two scenarios for adaptive capacity control.
To avoid stochastic variability, the scenario results were averaged from 20 model runs.

\textit{Pollution Scenarios}: Pollution scenarios consist of business as usual (BAU) and increase (INC). BAU refers to, ‘What if the seasonal pollution levels continue for another period?’, which assumes that the six-year time series of hourly PM\textsubscript{10} in 2010–2015 will replicate itself for another six years. INC projects an upward trend of the seasonal PM\textsubscript{10} averages by 3\% every season.

\textit{Resilience Scenarios}: The scenario controls the agent's maximum health to which recovery is possible. The variable is assigned as the ‘adaptive capacity (AC)’, and divided as AC100 and AC200, meaning that if the resilience is set for AC200 the agent whose health goes below a nominal value of 200 recovers to a maximum of 200.


\subsection{Scenario Results}
\textbf{BAU Scenario}: In the BAU×AC100 (see Figure~\ref{fig:bau_total}-Top), the `at-risk' population in most districts started to increase between 4500 and 6000 ticks, and from 7200 ticks to the end of the simulation. This is to be expected given that agents were given maximum health at the start of the simulation, and similar health loss occurred across agents in the same group as PM\textsubscript{10} generated from each district monitor gave similar exposure levels.

Furthermore, the time-series plots revealed an oscillating trend. This may have resulted from the balance between the exponentiality of health loss and the linearity of resilience from the resilience scenario and the land price effect. 
The districts with oscillating risk population trends were mostly in the AC100 scenario, with people aged 15 to 64. These groups appeared to be associated with the balance between recovery rate and health degrading with exposure. The districts that did not experience oscillation repetition had been continuously exposed to extreme PM\textsubscript{10} episodes, presenting health recovery ineffective. The oscillations only appeared in a few districts for the elderly group, but these districts experienced a linear upward trend due to the group's limited ability to recover from illness.


\begin{figure}[h]
 \begin{center} 
	\includegraphics[width=\textwidth]{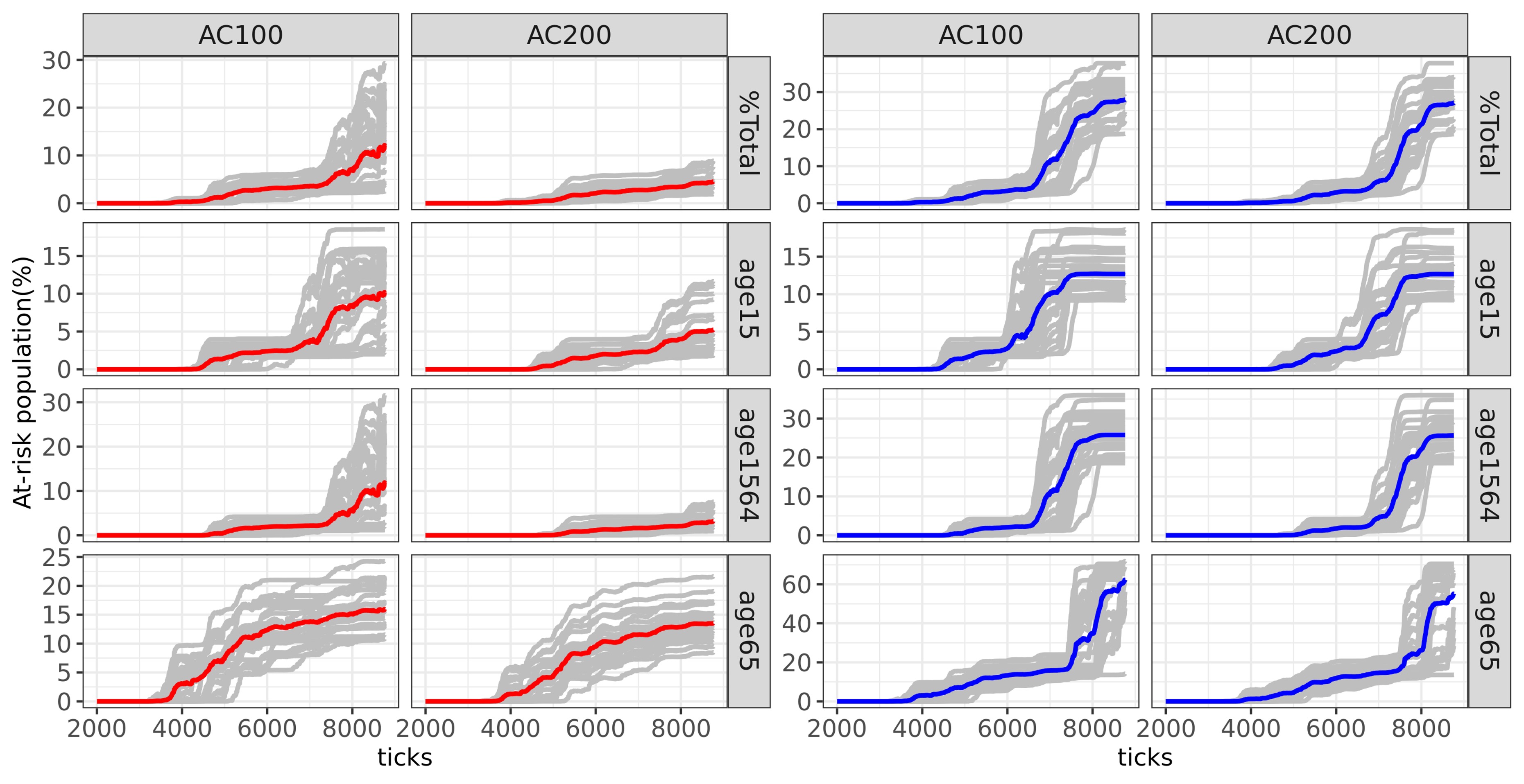}
 \end{center} 
 \vspace{-5mm}
 \caption{Population at risk from cumulative PM\textsubscript{10} exposure across 25 districts in BAU (Business as Usual) and INC (Increased Intervention) scenarios, post-onset. Variability in the at-risk population is evident in the final simulation step for both age groups under 15 and ages 15-64, with the BAU and INC scenarios displaying significant differences. Conversely, the variability for those aged 65 and above is less pronounced, indicating a more consistent impact across scenarios.} 
 \label{fig:scenario_result} 
 \end{figure} %

\begin{figure}[h]
 \begin{center} 
	\includegraphics[width=.8\textwidth]{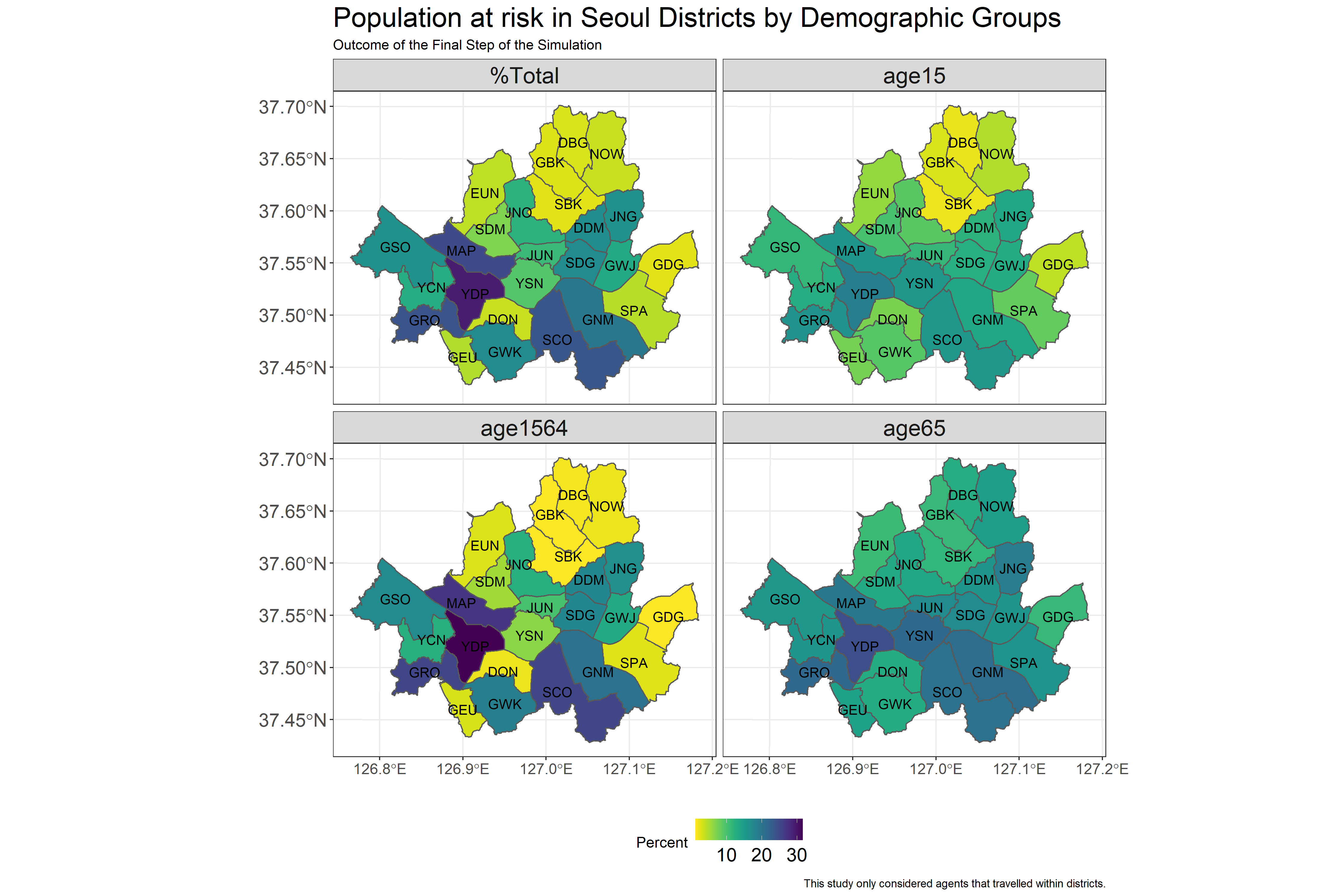}
 \end{center} 
 \vspace{-5mm}
 \caption{Health risk in Seoul Districts as a result of BAU$\times$AC100.} \label{fig:HV_Map} 
 \end{figure}

Using the outcome of BAU$\times$AC100 as a default structure, the highest risk areas included Yeongdungpo, Mapo, and Guro, which had over 20\% of the at-risk rate (see Figure~\ref{fig:HV_Map}). The total percentage was greatly affected by the 15-64 age group (`economically-active’) because the group accounted for the majority of the city’s population. 

\textbf{INC Scenario}: The `at-risk' population in the INC scenario showed a small rise around 5\% by 3500 ticks but showed a steep incline to 30\% after 7000 ticks (see Figure~\ref{fig:scenario_result}). Yeongdeungpo had the highest risk population at 37.8\% and the average risk population was 27.8\%. 

The temporal trend of the at-risk population varied by demographic group. While the over 65s had a constant increase until 7500 ticks with a final surge at around 75\%, the other two groups had a similar pattern that showed an initial onset at 4300 ticks, plateaued for a while, and surged in the final quarter of the simulation. 
Much of this occurred because the elderly lacked health resilience against the extreme PM\textsubscript{10} that was demonstrated in the BAU scenario outcome, whereas the other two groups had different times (ticks) of onset but had enough health resilience to maintain the at-risk level to some extent. 
However, the majority of the young and economically active people were able to keep their health from deteriorating due to the constant risk of PM\textsubscript{10}.

\section{Conclusion}\label{section:conclusion}
This study examines the impact of exposure levels and socioeconomic resilience on the health of Seoul's population, utilising an agent-based model. 
The study primarily analysed the sensitivity of health loss and additional pollution effect on road patches assuming people’s home and work locations are nearer to roads tended to trigger the risk.

The results from the scenario analysis indicate that vulnerability in all districts begins to manifest early, at 4000 ticks, and then significantly increases after 7000 ticks. 
This trend is closely linked to the point at which demographic groups fall into poor health (i.e., health below 100) and the extent of health recovery support available in each district. 
One possible explanation for this pattern is the model's initial assumption that all individuals start with a health status of 300, an oversimplification that does not account for the varying health conditions present in reality. 
The sharp increase in vulnerability is interpreted as a response to accumulated exposure to particulate matter. Therefore, it is suggested that future research should incorporate morbidity and, where possible, mortality statistics to provide a better understanding of the health impacts of pollution exposure.

Moreover, the study reveals that the vulnerability of the elderly and individuals with lower education levels rises significantly across different scenarios: over 10\% in the Business As Usual (BAU) scenarios, over 30\% in the Increased (INC) scenarios, and over 5\% in the Decreased (DEC) scenarios, regardless of the resilience strategies implemented.
These results build upon the work of Jerrett et al.\cite{Jerrett2001}, O'Neill et al.\cite{ONeill2003}, and Moreno-Jimenez \cite{Moreno-Jimenez2016}, highlighting the influence of long-term exposure to pollutants and geographical factors on health disparities, particularly among groups vulnerable to the effects of biological ageing. 
The study underscores the importance of considering the static nature of vulnerable populations, such as the elderly, who may lack access to adequate healthcare during pollution events. With a projected annual pollution increase of 3\%, the research indicates a sharp rise in the affected population after 7000 ticks (equivalent to 10 years), with initial resilience strategies against PM\textsubscript{10} showing minimal impact.

However, the model's reliance on a population proxy and the selection of pollution data for each location as a single value highlights its role as an \textit{'illustrative model'} rather than a direct replication of real-world scenarios \cite{edmonds2019}. It serves to elucidate the potential variations in health outcomes across different districts when exposed to pollution levels exceeding legal thresholds.

In conclusion, while providing valuable insights into the health effects of PM\textsubscript{10} exposure in Seoul at a mesoscale simulation, this paper calls for future research to explore other pollution thresholds and to develop more sophisticated models of population dynamics and pollution generation. The goal is to better capture the complex interplay between pollution exposure and health outcomes.

\section*{Data and Reproducible Code}
All data and executable codes are stored in the GitHub repository:

\noindent\url{https://github.com/dataandcrowd/PollutionABM}

%
%
%
\bibliographystyle{splncs04}
\bibliography{bibliography.bib}
\end{document}